\documentclass[11pt]{article}
\usepackage[numbers, sort&compress, square]{natbib}
\usepackage{dblfloatfix}
\usepackage{ulem}
\usepackage{caption}
\usepackage{dramatist}
\usepackage{xspace}
\usepackage{pifont} % http://ctan.org/pkg/pifont
\usepackage{multirow}
\usepackage{tcolorbox}
\usepackage{xltabular}
\usepackage{longtable}
\usepackage{hyperref}
\usepackage{listings}
\usepackage{amsfonts}
\usepackage{amsmath}
\usepackage{amssymb}
\usepackage{lineno}
\usepackage{multirow}
\usepackage{adjustbox}
\usepackage[english]{babel}
\usepackage[bottom]{footmisc}
\usepackage{booktabs}
\usepackage{CJKutf8}
\usepackage{subfigure}
\usepackage{setspace}

\usepackage{dsfont}
\usepackage{array} % 用于tabularx环境
\usepackage{tabularx} % 用于tabularx环境
\usepackage{subfigure} % 用于创建子图
\usepackage{xcolor} % 用于画有颜色的框

\usepackage{lipsum}  % 用于生成示例文本
\usepackage{multicol} % 引入 multicol 宏包
\usepackage{enumitem}
\usepackage{graphicx}
\usepackage{subcaption}
\usepackage{caption}

\usepackage{authblk}
\usepackage{geometry}
\lstdefinestyle{bashstyle}{
    language=bash,
    backgroundcolor=\color{gray!10},
    basicstyle=\ttfamily\small,
    frame=single,
    showstringspaces=false,
    columns=fullflexible,
    captionpos=b,
    keywordstyle=\color{blue},
    morekeywords={>>>}
}

\lstdefinestyle{pythonstyle}{
    language=Python,
    backgroundcolor=\color{gray!10},
    basicstyle=\ttfamily\small,
    frame=single,
    showstringspaces=false,
    columns=fullflexible,
    captionpos=b,
    keywordstyle=\color{blue},
    commentstyle=\color{gray},
    stringstyle=\color{teal},
    morekeywords={>>>}
}

\makeatletter
\def\@BTrule[#1]{%
  \ifx\longtable\undefined
    \let\@BTswitch\@BTnormal
  \else\ifx\hline\LT@hline
    \nobreak
    \let\@BTswitch\@BLTrule
  \else
     \let\@BTswitch\@BTnormal
  \fi\fi
  \global\@thisrulewidth=#1\relax
  \ifnum\@thisruleclass=\tw@\vskip\@aboverulesep\else
  \ifnum\@lastruleclass=\z@\vskip\@aboverulesep\else
  \ifnum\@lastruleclass=\@ne\vskip\doublerulesep\fi\fi\fi
  \@BTswitch}
\makeatother

\addto\extrasenglish{
}

 {\begin{list}{}%
         {\setlength{\leftmargin}{#1}}%
         \item[]%
 }
 {\end{list}}
 
\newif\ifrev
\revtrue
\ifrev
  \newcommand{\huifeng}[1]{{\color{cyan} [Huifeng: #1]}}
  \newcommand{\shijie}[1]{{\color{purple} [Shijie: #1]}}
  \newcommand{\yier}[1]{{\color{red} [Yier: #1]}}
\else
  \newcommand{\huifeng}[1]{}
  \newcommand{\shijie}[1]{}
  \newcommand{\yier}[1]{}
\fi

\newcommand{\projectname}{{CryptoTensors}}

\def\BibTeX{{\rm B\kern-.05em{\sc i\kern-.025em b}\kern-.08em
    T\kern-.1667em\lower.7ex\hbox{E}\kern-.125emX}}

\renewenvironment{abstract}{
  \smallskip
  \begin{center}
    {\Large\bfseries Abstract}
  \end{center}
  \vspace{-0.5em}
}
{}

\makeatletter
\renewcommand{\maketitle}{
    \begin{center}
        {\LARGE\bfseries \@title \par}
        \vspace{1em}
        {\@author}
    \end{center}
}
\makeatother

\title{\centering \projectname: A Light-Weight Large Language Model File Format for Highly-Secure Model Distribution}
\author{}  % 保持为空，不在 \author 里放 tabular
\date{}
\begin{document}
\begin{CJK*}{UTF8}{gbsn}

\maketitle   % 先画标题

% 紧跟在标题下面手动放作者两列表格
\begin{center}
\begin{tabular}{cc}
\begin{tabular}{c}
\textbf{Huifeng Zhu}\\
Huawei Technologies\\
\texttt{zhuhuifeng4@huawei.com}
\end{tabular}
&
\begin{tabular}{c}
\textbf{Shijie Li}\\
University of Science and Technology of China\\
\texttt{shijie\_li@mail.ustc.edu.cn}
\end{tabular}
\\[1.5em]
\begin{tabular}{c}
\textbf{Qinfeng Li}\\
Zhejiang University\\
\texttt{liqinfeng@zju.edu.cn}
\end{tabular}
&
\begin{tabular}{c}
\textbf{Yier Jin}\\
University of Science and Technology of China\\
\texttt{jinyier@ustc.edu.cn}
\end{tabular}
\end{tabular}
\end{center}

\vspace{2em}
\begin{abstract}
To enhance the performance of large language models (LLMs) in various domain-specific applications, sensitive data such as healthcare, law, and finance are being used to privately customize or fine-tune these models. Such privately adapted LLMs are regarded as either personal privacy assets or corporate intellectual property. Therefore, protecting model weights and maintaining strict confidentiality during deployment and distribution have become critically important. However, existing model formats and deployment frameworks provide little to no built-in support for confidentiality, access control, or secure integration with trusted hardware. Current methods for securing model deployment either rely on computationally expensive cryptographic techniques or tightly controlled private infrastructure. Although these approaches can be effective in specific scenarios, they are difficult and costly for widespread deployment.

In this paper, we introduce \projectname{}, a secure and format-compatible file structure for confidential LLM distribution. Built as an extension to the widely adopted Safetensors format, \projectname{} incorporates tensor-level encryption and embedded access control policies, while preserving critical features such as lazy loading and partial deserialization. It enables transparent decryption and automated key management, supporting flexible licensing and secure model execution with minimal overhead. We implement a proof-of-concept library, benchmark its performance across serialization and runtime scenarios, and validate its compatibility with existing inference frameworks, including Hugging Face Transformers and vLLM. Our results highlight \projectname{} as a light-weight, efficient, and developer-friendly solution for safeguarding LLM weights in real-world and widespread deployments.
\end{abstract}

% \begin{document}
% \begin{CJK*}{UTF8}{gbsn}

% \newpage

% \begin{spacing}{0.9}
% \tableofcontents
% \end{spacing}

\newpage

\section{Introduction}
The rapid and widespread adoption of large language models (LLMs) has led to a growing number of closed-weight models which cannot be publicly released due to privacy, commercial, or regulatory constraints. These models generally fall into two categories. The first comprises foundation models trained from scratch by major companies, such as GPT-5~\cite{achiam2023gpt}, Claude~\cite{anthropicclaude}, and Gemini~\cite{team2023gemini}. These models remain proprietary to safeguard substantial training investments and maintain competitive advantage. The second category includes domain-specific models fine-tuned from open-weight LLMs (e.g., LLaMA~\cite{touvron2023llama, metallama4}, Mistral~\cite{jiang2023mistral, jiang2024mixtral}, Gemma~\cite{team2024gemma}, DeepSeek~\cite{liu2024deepseek, guo2025deepseek}, Kimi~\cite{kimivl, kimik2}). The proliferation of open models has greatly lowered the barrier to entry, allowing developers to adapt them using proprietary or sensitive datasets in domains such as healthcare~\cite{nazi2024large}, law~\cite{shu2024lawllm}, and finance~\cite{lee2025large}. The resulting models often inherit the sensitivity of their training data and, consequently, cannot be openly shared.

Deploying these closed-weight models in broader settings exposes natural trade-offs between privacy, inference speed, and hardware cost~\cite{lin2025pushing}. Most current development efforts focus on improving performance, usability, and format compatibility—for example, optimizing inference efficiency~\cite{holmes2024deepspeed, shah2024flashattention} or introducing new model storage formats~\cite{ggufgithub, shah2025qstore}. Existing solutions for securely deploying proprietary LLMs, however, tend to be either too complex or inaccessible to most developers. In practice, large organizations depend on private cloud environments, custom communication protocols, encrypted delivery systems, or legal agreements to control model access~\cite{google_vertexai, aws_bedrock, ibm_watsonx}. While these strategies are effective at scale, they demand substantial infrastructure and legal resources, making them impractical for smaller teams or open model ecosystems. Conversely, academic research has explored privacy-preserving computation techniques that offer strong security guarantees. Yet, these approaches are difficult to apply in real-world AI workflows, as they often require model-specific modifications and heavy runtime support~\cite{rho2024encryption, de2024privacy}. As a result, there remains no lightweight, general-purpose solution that can provide secure and controlled model usage while remaining compatible with today’s common deployment pipelines.

Facing the above challenges, in this paper, we propose \projectname{}, a light-weight LLM file format for highly secure model distribution. To protect model weights, \projectname{} encrypts the model weights, ensuring that even if a user copies and obtains the entire model file, they cannot load or use the model properly. To minimize infrastructure requirements, \projectname{} only uses software-based encryption techniques. Most importantly, we achieve compatibility with Safetensors~\cite{Safetensors}, so our format leverages existing widely adopted models, ensuring compatibility with frameworks such as vLLM and Hugging Face Transformers. As a result, users only need a Python environment to decrypt the model weights and perform model inference. To ensure broad applicability, our encryption operates at the tensor level, encrypting each tensor (i.e., each weight matrix) with a unique data encryption key. This fine-grained encryption ensures that the encryption is independent of the model’s architecture, allowing it to be applied to any model structure, whether dense or sparse.
% Facing above challenges, in this paper, we propose \projectname{}, a LLM file format for highly secure model distribution. The proposed model file format provides a practical and lightweight entry point for protection: \textbf{\textit{compared to system-level encryption or custom deployment protocols, enhancing the format layer enables encryption and policy enforcement with minimal engineering effort and natural alignment with the model sharing workflow}}. Based on this insight, we chose to extend existing widely adopted formats rather than introducing an entirely new one, preserving compatibility with community tools and reducing migration overhead. Specifically, \projectname{} extends the popular Safetensors~\cite{Safetensors} format by introducing tensor-level encryption and embedded policy controls, while retaining key features such as lazy loading and partial deserialization. It remains fully compatible with existing model loading infrastructures like vLLM and Hugging Face Transformers. 
Overall, the primary contributions of this work are as follows:
\begin{itemize}
    % \item We introduce \projectname{}, a secure LLM file format that extends Safetensors, for closed-weight model controlled distribution and protection, providing strong confidentiality guarantees with practical overhead.\huifeng{Reduce the attack surface.}
    \item We introduce \projectname{}, a secure LLM file format that extends Safetensors for controlled distribution and protection of closed-weight models. Following standard security practice, we minimize the trusted computing base-the subset of software and hardware components that must be assumed trustworthy-to a small set of cryptographic primitives and avoid system-level dependencies, thereby reducing the attack surface and ensuring strong confidentiality guarantees with practical overhead.
    \item \projectname{} adopts a decoupled structure where the file header is in plaintext but its integrity is protected through digital signatures. The tensor body is protected via fine-grained encryption. This enables compatibility with existing indexing mechanisms and preserves essential optimizations such as lazy loading and partial deserialization.
    \item The file format is self-contained, embedding cryptographic metadata to enable transparent and policy-compliant decryption for authorized users without manual intervention. Our design supports automated key provisioning and policy validation through integration with remote Key Broker Services (KBS), enabling flexible licensing models and secure deployments across diverse environments.
    \item We implement a proof-of-concept library and comprehensively benchmark its performance, including serialization/deserialization overhead and runtime characteristics. We further demonstrate its practical applicability by integrating it into real-world inference frameworks such as Hugging Face Transformers and vLLM.
\end{itemize}

\section{Background}
\subsection{Related Work}
% \begin{itemize}
%     \item \textbf{Confidentiality}: Preventing unauthorized access to the model’s internal parameters. This is crucial for protecting proprietary or sensitive models from being copied, reverse-engineered, or leaked.
%     \item \textbf{Integrity}: Ensuring that the model has not been tampered with or corrupted since it was serialized. An attacker who modifies model weights can introduce harmful biases or backdoors, potentially leading to catastrophic failures in deployment.
% \end{itemize}

% Current secure formats like Safetensors do not natively support either of these properties. Once a .safetensors file is obtained, any user can inspect or alter its contents without detection. This creates a significant gap in the security model for organizations that rely on these formats to protect high-value intellectual property.

% Within the paradigm of LLMs, data has emerged as the fundamental and most strategically valuable asset, necessitating comprehensive security frameworks to ensure robust protection across the entire lifecycle. 
Protecting the weights of LLMs has become an increasingly important concern in both academia and industry. Given their substantial training costs and the proprietary data involved, model weights are valuable intellectual assets but are vulnerable to various security threats, including model stealing~\cite{mukherjee2023orca, carlini2024stealing, liu2025model}, unauthorized access~\cite{li2025commercial}, and illegal redistribution~\cite{ren2024copyright}.
% Nevo et al.~\cite{nevo2024securing} proposed 38 different attack vectors against model weights security and defined 5 security levels for protecting model weights.

To address these challenges, researchers have proposed several lines of defense.
% One common approach is model watermarking, which embeds hidden patterns or behaviors into the model for the purpose of ownership verification. Watermarking methods can be black-box (detectable via specific prompts)~\cite{kirchenbauer2023watermark, pan2024markllm} or white-box (embedded in the model weights or activations)~\cite{bahri2024watermark, chang2024postmark}, each with different trade-offs in robustness and stealth. However, watermarking serves only as a passive detection mechanism for model theft and cannot actively prevent it. 
One popular method is model watermarking, which embeds identifiable patterns either in model behavior (black box)~\cite{kirchenbauer2023watermark, pan2024markllm} or in internal representations such as weights or activations (white box)~\cite{bahri2024watermark, chang2024postmark}, enabling ownership verification. However, watermarking is inherently passive-it may help detect theft but cannot prevent unauthorized use or redistribution.

Privacy-preserving machine learning (PPML) seeks stronger guarantees by protecting models and data during execution. Homomorphic encryption (HE) allows computations to be performed directly on encrypted inputs, safeguarding both model weights and user data~\cite{acar2018survey}. HE has been applied in secure neural inference~\cite{gilad2016cryptonets}, encrypted training~\cite{HELeTrieu, PrivTuner}, and federated learning settings where client gradients are encrypted before aggregation~\cite{jiang2021flashe,zhang2020batchcrypt,hu2024maskcrypt}. Despite its effectiveness, HE incurs substantial computational overhead and lacks efficient support for nonlinear operations such as GELU, softmax, or LayerNorm~\cite{rho2024encryption}. These limitations make HE difficult to scale for large models. Even the fully encrypted inference of a model with about 100M parameters requires several minutes per forward pass~\cite{de2024privacy}, making HE impractical for real-world deployment at scale.

Secure multi-party computation (SMPC) offers another direction by enabling distributed computation without revealing private inputs. It builds on primitives such as secret sharing~\cite{shamir1979share}, garbled circuits~\cite{yao1986generate}, and oblivious transfer~\cite{rabin2005exchange}. Frameworks like SecureML~\cite{mohassel2017secureml} and Marill~\cite{rathee2024mpc} use SMPC to enable collaborative training and secure LLM inference. Nonetheless, SMPC remains difficult to scale to large LLMs due to its high communication and computational cost~\cite{fan2022nfgen}.

In the industry, protection is often handled through private deployment pipelines and encryption integrated into proprietary runtimes. OpenVINO~\cite{openvinoOpenVINOx2122Security} and Windows ML~\cite{microsoftEncryptedModels} support encrypted model execution, but are restricted to specific formats (e.g., IR, ONNX) and require close coupling with proprietary runtimes. CoreWeave’s Tensorizer~\cite{tensorizer} enables chunked encryption for fast cloud-based loading, but relies on a custom format with limited key management and is tailored for serverless scenarios-limiting its portability to other workflows. In broader practice, many commercial deployments depend on private-cloud inference or legal agreements to control distribution. These strategies can be effective in enterprise settings but are inflexible, non-standardized, and largely inaccessible to small teams or individual developers.

In contrast to these heavyweight cryptographic approaches and tightly coupled industrial framework, academic research has proposed privacy-preserving techniques such as TEE-based inference~\cite{li2024coreguard} and GPU-accelerated FHE-based inference~\cite{de2024privacy}. While these methods offer strong security guarantees, their applicability in real-world scenarios remains limited. These solutions often require significant computational overhead, specialized infrastructure, or are highly tailored to specific use cases, making them difficult to integrate into mainstream AI workflows. Moreover, they tend to be complex to implement and require extensive runtime support, limiting their adoption outside of specific academic environments.

Our work aims to provide strong model protection with minimal disruption to existing infrastructure. By extending widely adopted formats and preserving compatibility with frameworks such as Hugging Face and vLLM, we enable scalable, efficient, and accessible protection that bridges the gap between academic security mechanisms and real-world usability.
% In addition, access control mechanisms are widely used in practical systems for model weight security. Xiao et al.~\cite{xiao2025privacy} proposed a framework based on attribute-based encryption (ABE) to manage an LLM-driven electrical distributed systems, achieving fine-grained access control on LLM.

% THEMIS~\cite{themis} is proposed to enhance the security and trustworthiness of compiled ML models. To address concerns about supply chain attacks and intellectual property leakage in MCaaS, THEMIS integrates property-based integrity using a "model property tree" for transparent yet privacy-preserving verification of the compilation process. Furthermore, it enforces fine-grained confidentiality and access control over model binaries through ABE-based hybrid selective encryption. The entire service is secured within distributed Trusted Execution Environments (TEEs) to protect against advanced threats and ensure reliable execution. \shijie{move to related work}

\subsection{Safetensors}

\projectname{} builds upon Safetensors~\cite{Safetensors}, a model serialization format introduced by Hugging Face as a secure alternative to PyTorch’s Pickle-based weight files~\cite{safetensors_github}. Traditional Pickle-based formats suffer from critical vulnerabilities~\cite{slaviero2011sour, hiddenlayerHijackingSafetensors}. That is, they allow arbitrary code execution during deserialization, posing a major security risk. Safetensors was specifically designed to eliminate this risk by using a pure data format without any embedded executable logic.

\begin{figure}[h]
    \centering
    \includegraphics[width=\linewidth]{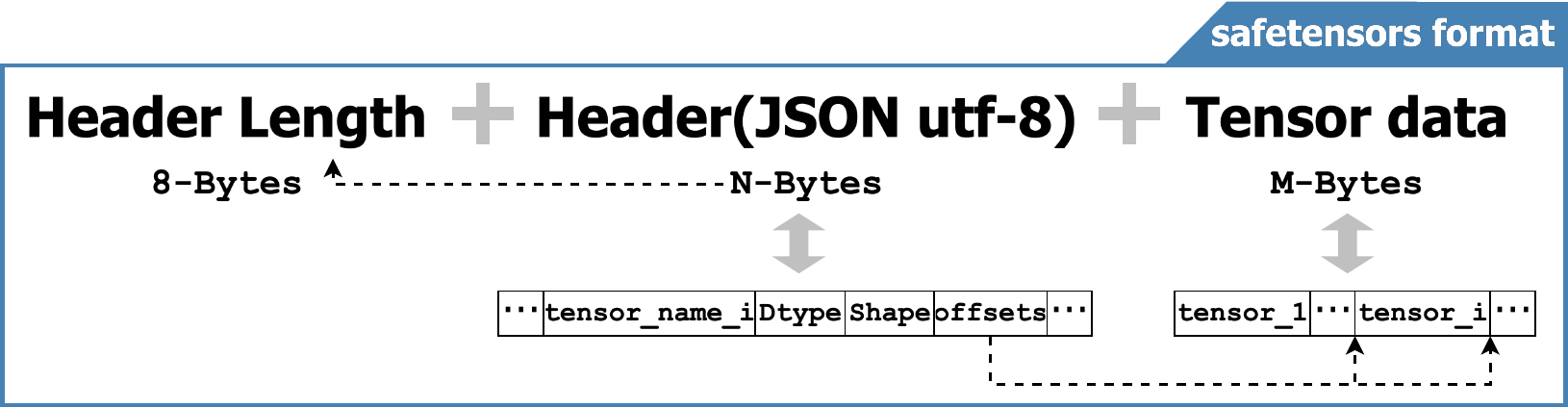}
    \caption{Structure of a Safetensors File.}
    \label{fig:safetensors}
\end{figure}

Safetensors has since become one of the most popular formats for publishing and loading large language models (LLMs), with native support in major inference frameworks such as Hugging Face Transformers~\cite{wolf2020transformers}, vLLM~\cite{kwon2023efficient}, etc. Many state-of-the-art models~\cite{liu2024deepseek, guo2025deepseek, metallama4, yang2025qwen3} are now released with Safetensors as the default format.

As illustrated in Figure~\ref{fig:safetensors}, a Safetensors file consists of three segments: an optional N-byte prefix, a JSON-formatted header, and a raw tensor data section (the body). The header describes the metadata of each tensor-such as its name, shape, data type, and byte offsets-while the body contains the corresponding binary tensor payloads. This clean separation between metadata and data enables powerful optimizations.

One of the most important design features enabled by this structure is lazy loading. Since the file header encodes precise offsets for each tensor, inference engines can parse only the header and then load individual tensors on demand. This allows large models to be partially loaded into memory, greatly reducing peak usage and improving performance in use cases like early-exit models or multi-stage pipelines. Moreover, tensors in the body are stored in a flat binary layout, making them amenable to direct memory mapping (mmap), which enables zero-copy deserialization.

Several performance-optimized implementations have leveraged these properties. For example, fastsafetensors~\cite{yoshimura2025speeding} achieves accelerated loading by exploiting asynchronous I/O and parallel tensor mapping, made possible only by Safetensors’ decoupled structure. Similarly, other emerging model formats such as GGUF~\cite{ggufgithub} also adopt similar features of decoupling metadata and data to support fast and modular loading workflows.

However, conventional encryption approaches are often incompatible with this file format. Encrypting the entire model file, including the header, breaks compatibility with existing tooling and disables key features like lazy loading and memory mapping. The header must remain readable to enable indexing, shape inference, and tensor slicing, all of which are critical for efficient inference. Once encrypted, these capabilities are lost, and the model must be fully decrypted before use, resulting in significant overhead and integration complexity.

\section{Threat Model}

%In this paper, we try to address the critical issue in making proprietary LLM models transactional. That is, we will develop a proprietary model trading platform, with the support of \projectname{} file type, such that model owners can easily and safely trade their models. 
%We focus on protecting the confidentiality and integrity of the model weights. Our goal is to ensure that, even if the encrypted model file is leaked or shared publicly, it remains unusable and unintelligible to unauthorized parties. 
%Our threat model assumes an adversary who has full access to the model file. This includes the ability to inspect, copy, and attempt to reverse engineer the file using any available tools. We assume that a state-of-the-art key management system (KMS) is available and is not broken (note that all cloud servers provide such services). The attacker does not possess the decryption key and cannot access any KMS or trusted execution environment (TEE) where the key may be securely stored. Without the correct decryption key, it should be computationally infeasible for an adversary to recover meaningful model content or use the model in any way. 

In this paper, we try to address the critical issue of proprietary models protection. More specifically, we focus on protecting the confidentiality and integrity of the model weights. 
Our threat model assumes an adversary who has full access to the model file but does not have access to the decryption keys of the tensors. 
This includes the ability to inspect, copy, and attempt to reverse engineer the file using any available tools. We assume that a state-of-the-art KBS is available and can be trusted. 
Without the correct decryption key, it should be computationally infeasible for an adversary to recover meaningful model content or use the model in any way. 
The protection of KBS as well as the key distribution system, is out of the scope of this paper. 

\section{\projectname}
\begin{figure}[!t]
\centering
\includegraphics[width=0.99\textwidth]{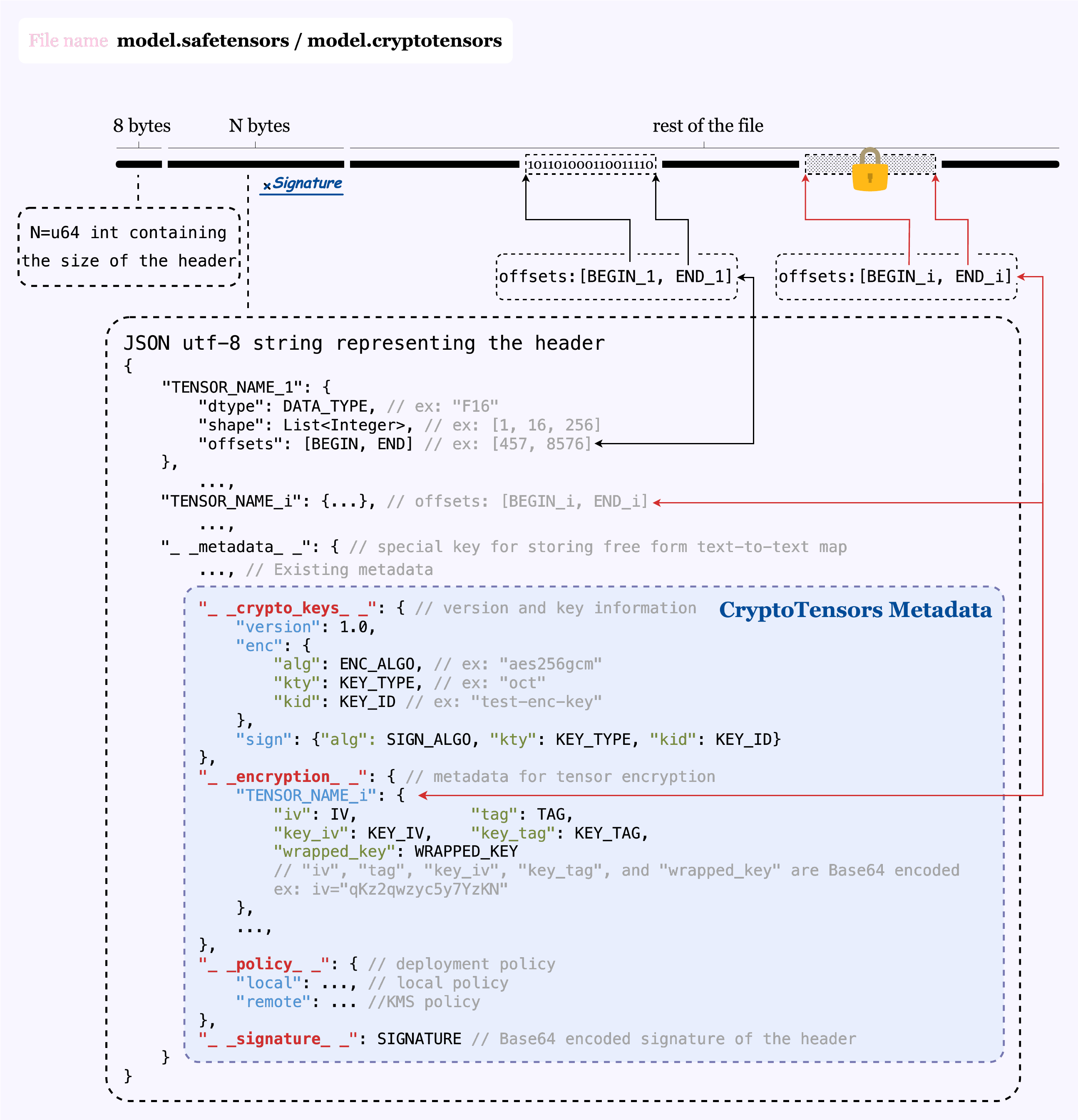}
\caption{
    The file format design of \projectname{}, extended from Safetensors. The blue components highlight the additional fields introduced to support tensor encryption, policy enforcement, and header integrity.
}
\label{fig:cryptotensors}
\end{figure}

In this section, we present a detailed description of \projectname{}, covering its design principles, file format structure, end-to-end serialization and deserialization workflow, and our proof-of-concept implementation.

\subsection{Design Principles}

\projectname{} is designed to provide strong confidentiality and controlled access for proprietary LLMs while maintaining compatibility with existing machine learning ecosystems. Rather than introducing a new format from scratch, \projectname{} extends the widely adopted Safetensors format to support encryption and policy enforcement. This ensures seamless integration with current tooling-such as Hugging Face Transformers and vLLM-while minimizing migration overhead. By building on a format already optimized for inference efficiency, \projectname{} inherits features like lazy loading and partial deserialization, which are critical for handling large models.

A central design principle of \projectname{} is self-containment. The model file embeds all necessary cryptographic metadata, including encrypted tensor keys, policy references, and key source descriptors. This allows a compliant inference engine or model loader to retrieve decryption keys securely from a remote KBS without relying on external metadata services. All required validation and policy checks can be performed using the information within the file and the runtime environment, reducing deployment complexity and eliminating the need for manual key handling or external configuration.

The file structure of \projectname{} preserves the header-body separation of Safetensors. The header remains unencrypted to allow indexing systems, loaders, and model hubs to inspect non-sensitive metadata such as tensor names, shapes, and data types. The proposed format enables the continued use of performance optimizations and model discovery mechanisms that depend on header visibility. To prevent tampering or unauthorized modification, the header is protected by a digital signature, which covers all metadata and offsets and ensures file integrity and authenticity prior to decryption.

For secure tensor storage, \projectname{} adopts a two-level encryption strategy. Each tensor is encrypted independently using a randomly generated data encryption key (DEK) and a unique initialization vector (IV), ensuring ciphertext uniqueness and mitigating the risks of key or IV reuse. These DEKs are then encrypted (wrapped) using a master key governed by embedded access policies. This design enables fine-grained access control, reduces attack surfaces, and supports efficient partial decryption. Combined with \projectname{}'s self-contained structure and policy-aware key retrieval, the format provides a lightweight and secure foundation for distributing closed-weight models across diverse research and production environments.

\subsection{\projectname{} Format}

The format of \projectname{}, as shown in Figure~\ref{fig:cryptotensors}, is designed to be as simple as possible while supporting the flexibility required for secure and scalable model distribution. It extends the Safetensors format, which already includes a \colorbox{gray!20}{\texttt{\_\_metadata\_\_}} field to accommodate structured extensions. \projectname{} introduces four additional metadata fields, all of which are stored in this section and protected by a digital signature to guarantee integrity. These four fields are listed as follows.

\noindent\textbf{\textit{\_\_crypto\_keys\_\_:}} this field contains metadata related to encryption (\colorbox{gray!20}{\texttt{enc}}) and signing keys (\colorbox{gray!20}{\texttt{sign}}). This information is structured using the standard JSON Web Key (JWK) format~\cite{rfc7517} and may include attributes such as the key identifier (kid), URI-based key references (jku), and X.509 certificate chains (x5c). These references support flexible key retrieval mechanisms, including local files (file://), HTTP(S) endpoints, or secure key brokers (kbs://). The field also includes version information (\colorbox{gray!20}{\texttt{version}}) to ensure compatibility across future evolutions of the format.

\noindent\textbf{\textit{\_\_encryption\_\_:}} this field encodes all cryptographic data necessary to decrypt the tensors in the file. \projectname{} supports partial encryption, allowing users to encrypt only specific tensors. This flexibility enables a trade-off between performance and security, as only sensitive tensors incur the overhead of encryption and decryption. 
For each encrypted tensor, identified by its name and corresponding index in the header, this field stores the associated encryption metadata. Specifically, it includes the IV (\colorbox{gray!20}{\texttt{iv}}) and authentication tag (\colorbox{gray!20}{\texttt{tag}}) used to decrypt the tensor data with the DEK, as well as the wrapped DEK itself (\colorbox{gray!20}{\texttt{wrapped\_key}}), along with the IV (\colorbox{gray!20}{\texttt{key\_iv}}) and tag (\colorbox{gray!20}{\texttt{key\_tag}}) used to decrypt the DEK with the master key. All of this data is Base64-encoded to ensure portability and compatibility with standard transport and serialization formats.

\noindent\textbf{\textit{\_\_policy\_\_:}} this field defines a set of access control policies that govern both local model loading and remote key retrieval. Policies are expressed using Rego~\cite{Rego}, a declarative policy language supported by the Open Policy Agent framework~\cite{opa}. This field is subdivided into two sections: the local policy (\colorbox{gray!20}{\texttt{local}}), which is evaluated by the model loader to determine whether the execution environment meets the required conditions; and the remote policy (\colorbox{gray!20}{\texttt{remote}}), which is enforced by the KBS to decide whether the decryption key can be released. Embedding these policies directly in the file enables a high degree of portability and autonomy in policy enforcement, without reliance on external configurations.

\noindent\textbf{\textit{\_\_signature\_\_:}} this field contains a Base64-encoded digital signature of the file header. This signature protects the header from tampering and confirms its authenticity. It is verified both by the model loader-ensuring that the encryption metadata has not been altered-and by the KBS-validating that the embedded policies are trustworthy and unmodified. The presence of this signature establishes trust in both the structure and origin of the file, which is critical for secure deployment.

% \huifeng{Need a flowchart figure here.}
\begin{figure}[h]
  \centering
  \includegraphics[width=0.9\linewidth]{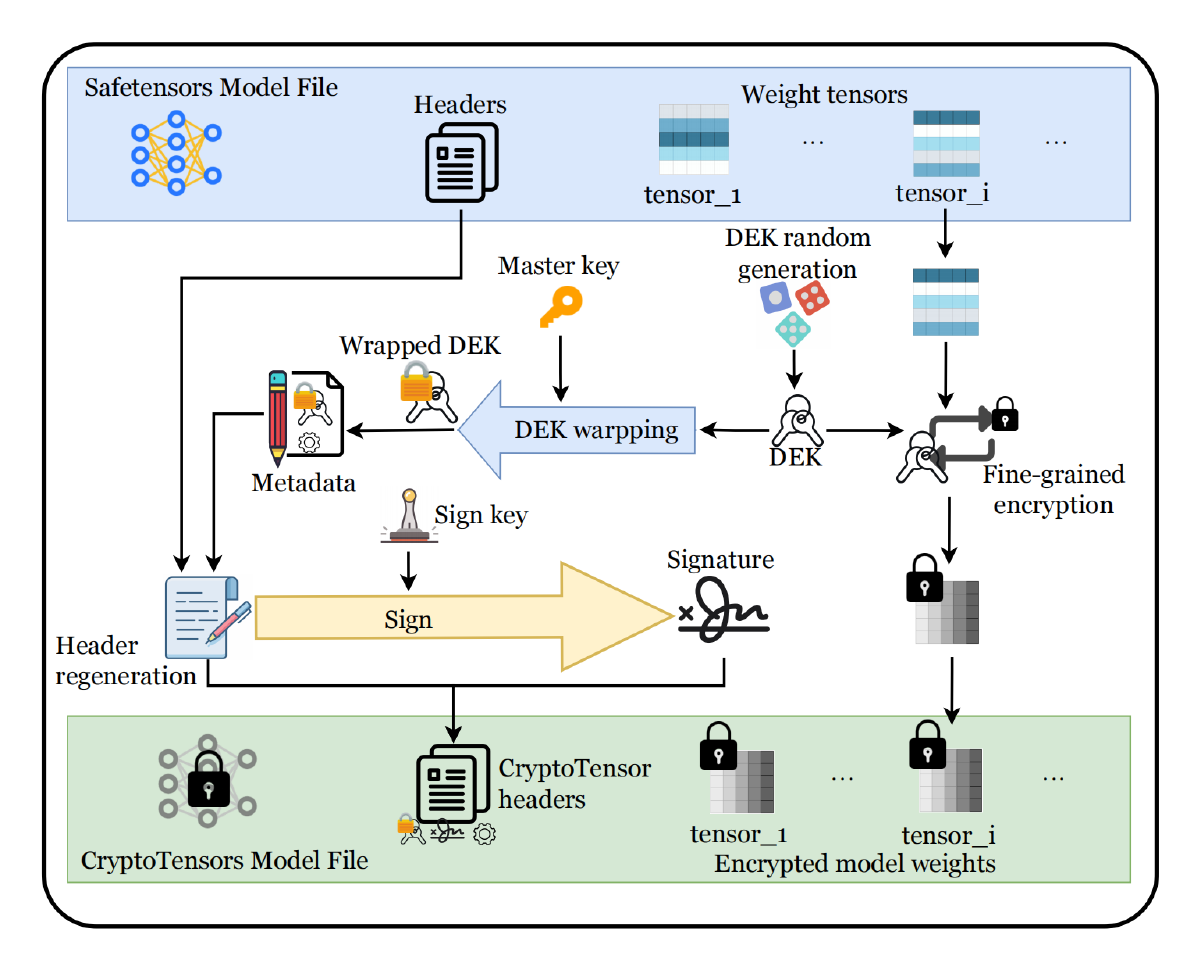}
  \vspace{-4mm}
  \caption{The process of serializing data into the \projectname{} format.}
  \label{fig:serialization}
\end{figure}

\subsection{Serialization to \projectname{}}
The procedure for serializing data into the \projectname{} format is illustrated in Figure~\ref{fig:serialization}.

Serialization involves three key steps: encrypting selected tensors, signing the file header, and embedding all necessary cryptographic metadata. Prior to serialization, the user specifies the encryption algorithm (e.g., AES-GCM-256), the corresponding master key, a signature algorithm (e.g., Ed25519) with its associated public/private key pair, a list of tensor names to be encrypted, and the deployment policies.

For each tensor selected for encryption, a unique data encryption key (DEK) is randomly generated. The tensor data is then encrypted using this DEK. To protect the DEK itself, it is wrapped using the master key. The encrypted tensor data and associated metadata are then prepared for serialization.

All cryptographic metadata, including key source information, encryption metadata per tensor, and access control policies, are stored in the \colorbox{gray!20}{\texttt{\_\_metadata\_\_}} section of the file header. This collection of metadata is assembled into a temporary header, which is then digitally signed using the provided private key. The signature is also included in the \colorbox{gray!20}{\texttt{\_\_metadata\_\_}} field, ensuring the integrity and authenticity of the header content.

Encrypted tensor data is written directly to the output file. Importantly, because initialization vectors (IVs) and authentication tags are stored separately from the data, the memory layout of encrypted tensors is identical to that of unencrypted tensors. This design avoids the need to recompute or shift tensor offsets during loading, ensuring that \projectname{} remains structurally compatible with existing loading mechanisms. This compatibility eliminates the need for offset recalculation or format conversion, ensuring that \projectname{} can be loaded using existing Safetensors-compatible tooling. In fact, a loader that supports Safetensors can directly load a \projectname{} file, with the only difference being that the tensor data will be encrypted.

Once serialization is complete, the resulting \projectname{} file can be safely distributed, including via public model hubs, without compromising model confidentiality. The master key, however, must be distributed securely only to authorized users, as access to the key is required for tensor decryption and model execution.

\subsection{Deserialization from \projectname{}}

\begin{figure}[h]
  \centering
  \includegraphics[width=0.9\linewidth]{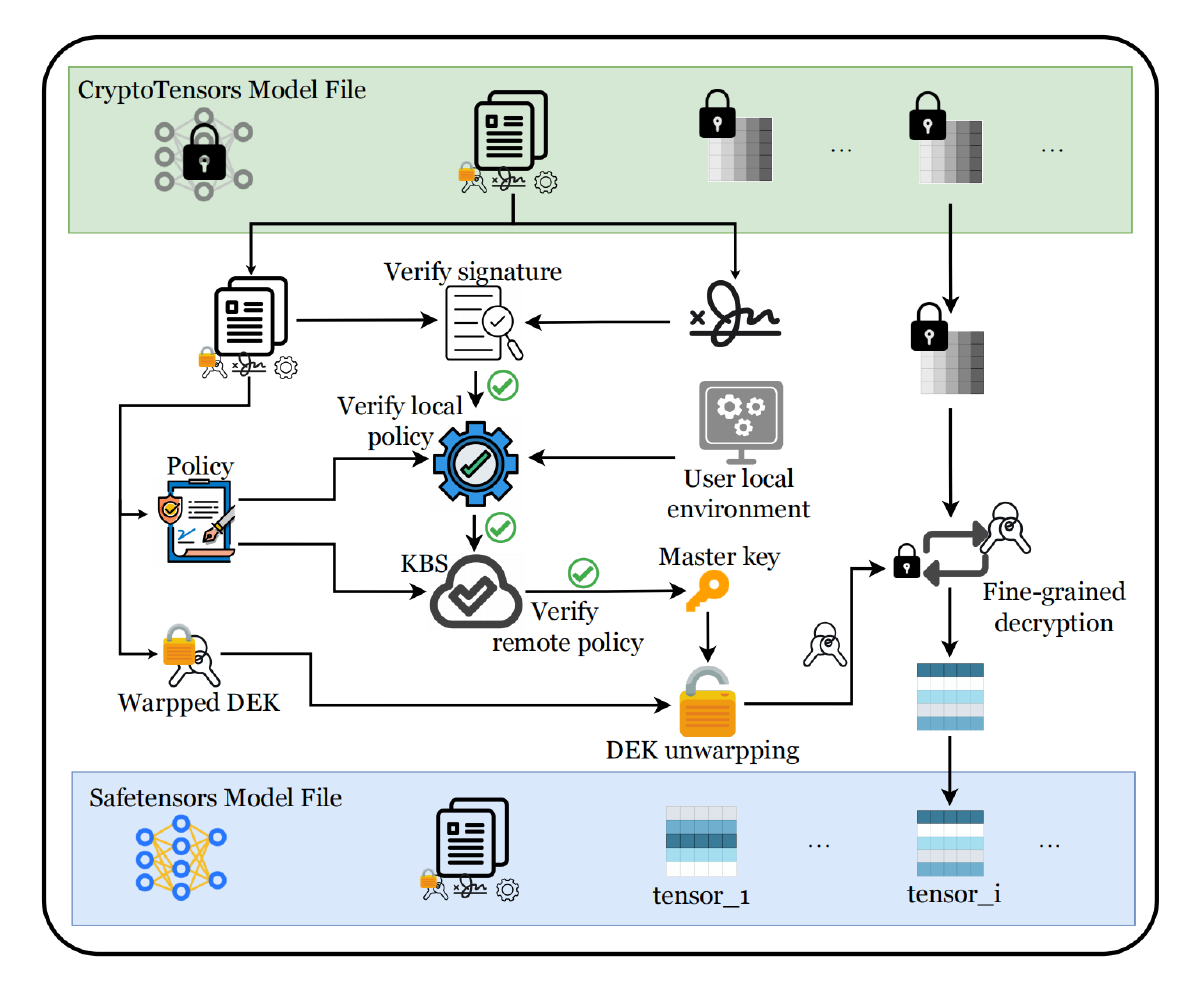}
  \vspace{-4mm}
  \caption{The process of deserializing data from the \projectname{} format.}
  \label{fig:deserialization}
\end{figure}

As shown in Figure \ref{fig:deserialization}, once a \projectname{} file is downloaded to the local filesystem, the model loader initiates the deserialization process. This process begins by parsing the file header to determine whether the file is encrypted. Specifically, the loader checks for the presence of the \colorbox{gray!20}{\texttt{\_\_crypto\_keys\_\_}} field. If this field is absent, the file is treated as a standard Safetensors file. If the field is present, the loader proceeds with the decryption workflow defined as below.

% \clearpage  

For encrypted files, the loader will first retrieve the signing key using the source information embedded in the header. % (e.g., \texttt{https://kbs.com}). 
\projectname{} is designed to make key retrieval fully automatic and transparent, enabling seamless access whether keys are stored locally or provided via remote services such as KBS. Once the signing key is retrieved, the loader verifies the integrity of the header by reassembling a temporary version of the header (excluding the signature field) and validating it against the signature included in the model file.

Upon successful signature verification, the loader evaluates the local deployment policy specified in the \colorbox{gray!20}{\texttt{\_\_policy\_\_}} section. This may involve collecting system-level measurements, such as device identifiers or attestation reports, depending on the conditions defined in the policy. If the local policy is satisfied, the loader proceeds to retrieve the master encryption key, again using the key source metadata embedded in the file.

If the master key must be obtained from a remote source, the loader establishes a secure connection and transmits the header along with the relevant measurements. The remote system then verifies the header’s integrity and evaluates the deployment policy against the provided measurements. If the policy check passes, the master key is returned, allowing decryption to proceed.

% All cryptographic keys in \projectname{} are represented using the JWK format to ensure standardization and interoperability. The loader should supports JWK Sets, enabling the inclusion of multiple keys for redundancy or key rotation. Key retrieval mechanisms are designed to be flexible and robust. The loader first examines the jku or similar fields provided in the header. If these fields are missing or inaccessible, the loader will fall back to checking environment variables. If neither is present, a default local path is used as a final fallback.\yier{Similar to last paragraph. Shall we keep it?}

To optimize performance and memory efficiency, \projectname{} supports lazy decryption in conjunction with lazy loading. Only tensors that are explicitly accessed during runtime are decrypted and loaded into memory. This approach is particularly beneficial for large models where only a subset of tensors may be required for inference or fine-tuning, significantly reducing both loading time and memory footprint.

\subsection{Proof-of-Concept Implementation}

We implement a PoC version of \projectname{}, which is a Python library, to demonstrate the feasibility of our format design and evaluate its performance in practice. The core logic is written in Rust, following the architecture of the original Safetensors library to maintain compatibility and ensure low-level memory safety and high throughput. On top of the Rust implementation, we expose Python bindings, enabling easy integration with mainstream machine learning frameworks such as Hugging Face Transformers and vLLM.

The \projectname{} library supports two underlying storage mechanisms: standard memory-mapped files (mmap) and PyTorch’s UntypedStorage. These backends serve as the foundation for integration with higher-level frameworks. Specifically, mmap is used to support NumPy, which in turn enables compatibility with a variety of downstream frameworks that rely on NumPy arrays. PyTorch integration is handled directly via UntypedStorage. This design ensures that \projectname{} remains interoperable across diverse machine learning ecosystems while maintaining high performance and minimal memory overhead.

While both storage mechanisms rely on memory mapping, their runtime behavior diverges significantly in practice. NumPy leverages standard OS-level mmap, whereas PyTorch uses its own UntypedStorage abstraction for managing tensor memory. In our implementation, we observed that accessing PyTorch’s storage layer from Rust introduces performance penalties due to the overhead of bridging with Python’s dynamic runtime. To mitigate this, encrypted files are handled directly via mmap within Rust, enabling efficient decryption and fast transfer of tensor data-potentially to accelerators memory-without unnecessary overhead.

To avoid redundant cryptographic overhead, the data will be decrypted once, only upon its first access during model loading. \projectname{} performs decryption and caches the resulting plaintext tensor in memory. Subsequent accesses to the same tensor bypass the decryption step entirely. %, ensuring that each tensor is decrypted only once. 

% To maximize performance, all encryption and decryption operations are implemented natively in Rust and designed to leverage parallel processing. When tensor data exceeds a given size threshold, multiple threads operate in parallel to encrypt or decrypt the data, exploiting the inherent parallelism of algorithms such as AES-GCM. For APIs like load\_file and save\_file, where all tensors are read or written, multi-threading is used across tensors, with each thread handling a different subset. We further optimize cryptographic operations by incorporating hardware-accelerated instructions when available, resulting in significant performance gains.

In terms of interface, \projectname{} is API-compatible with the existing Safetensors library. The serialization API accepts an optional configuration argument (as listed in Listing~\ref{lst:cryptotensors-save}); when this argument is provided, the data is encrypted and stored as a \projectname{} file. If omitted, the file is saved in the standard Safetensors format. For deserialization, the API remains unchanged. Thanks to transparent key retrieval and decryption, users do not need to modify the inference framework when loading encrypted files. This design ensures a high degree of backward compatibility and simplifies the adoption of \projectname{} in existing pipelines.

\vspace{1em}
\begin{lstlisting}[style=pythonstyle,
    caption={Saving an encrypted model with \projectname{} by passing a configuration argument},
    label={lst:cryptotensors-save}]
>>> from cryptotensors.torch import save_file
>>> save_file(model, model_out, config=config, metadata=metadata)
\end{lstlisting}

% To further ease integration, we provide a drop-in adapter package for transitioning from Safetensors to \projectname{}. Users first manually uninstall the original safetensors library and then install a compatibility layer by running the commands shown in Listing~\ref{lst:cryptotensors-migrate}. This adapter installs both the \projectname{} core library and a wrapper module that mimics the safetensors API. All calls to the safetensors module are internally redirected to the \projectname{} implementation.
% As a result, existing applications can continue using the original API without modification, while transparently gaining support for encryption and secure policy enforcement. The code remains agnostic to whether a file is encrypted or not, making the migration process virtually zero-effort. \yier{This paragraph may be removed since it is about real-world implementation practice, not related to the paper.}

% \vspace{1em}
% \begin{lstlisting}[style=pythonstyle,
%     caption={Migrating from Safetensors to \projectname{} via compatibility package},
%     label={lst:cryptotensors-migrate}]
% >>> pip uninstall safetensors
% >>> pip install cryptotensors-compatible
% \end{lstlisting}

% \huifeng{The data will be decrypted once}

\section{Evaluation}
% \section{Evaluation}

In this section, we systematically evaluate the performance and practicality of our proof-of-concept implementation to validate the design of the \projectname{} file format. The experiments are organized into three parts: we first measure serialization and deserialization overhead under various model sizes and frameworks, then analyze the trade-offs introduced by partial encryption and lazy decryption, and finally conduct end-to-end inference tests on mainstream engines such as Transformers and vLLM. The results demonstrate that \projectname{} introduces only moderate overhead while preserving full compatibility with existing model development and deployment workflows.

\subsection{Experimental Settings}

\noindent\textbf{\textit{Models.}}  
To cover a range of model sizes, we evaluate six LLMs from the Qwen3 family, spanning from 0.6B to 32B parameters.
These models are representative of both lightweight and large-scale deployment scenarios. All experiments are performed on a server equipped with a commercial %an Intel(R) Xeon(R) Silver 4310 
CPU and 512 GB of RAM. Model files are read and written to a 10TB 7200 RPM HDD, % (Western Digital WD101EFBX‑68B0AN0), 
providing realistic I/O behavior for large-scale models. Each benchmark is repeated 20 times, and we report the arithmetic mean.

\noindent\textbf{\textit{Implementation Details.}}  
We conduct experiments using two representative frameworks: PyTorch and NumPy. PyTorch is widely used across both academia and industry for training and inference, while NumPy serves as the foundational array library underlying many other frameworks such as TensorFlow, JAX, and Paddle. This choice ensures that our evaluation reflects the performance impact of \projectname{} library across a broad spectrum of real-world ML infrastructure. Both frameworks support Safetensors, enabling a direct comparison under consistent conditions.

\noindent\textbf{\textit{Baselines.}}  
To isolate the overhead introduced by \projectname{} library, we evaluate three configurations: (1) \textit{Safetensors}, the baseline Safetensors implementation loading standard unencrypted model files; (2)\textit{\projectname{}-Unencrypted},  \projectname{} implementation loading unencrypted model files; (3) \textit{\projectname{}-Encrypted}, implementation loading fully encrypted model files using AES-GCM-256 encryption and ED25519 signatures.
This setup enables us to separately quantify compatibility overhead and encryption-related costs. Unless otherwise noted, timing results are obtained using Python’s \texttt{time} module, and peak memory usage is measured with the \texttt{memory\_profiler} library. This experimental configuration is used consistently across all benchmarks.

% We focus on the core performance metrics related to serialization and deserialization. In particular, we evaluate the additional time and memory overhead introduced by our encryption framework. 

\subsection{Overhead of Serialization}

For serialization overhead benchmarking, we measure two core metrics: the time required to serialize all model parameters, and the peak memory consumption during the write process. We evaluate two serialization interfaces provided by our implementation: \texttt{save\_file}, which writes directly to disk, and \texttt{save}, which returns a serialized in-memory buffer. For clarity, we refer to the latter as \texttt{save\_bytes} in the remainder of this section.
For completeness, we also evaluate the file size overhead introduced by the embedded cryptographic metadata.

\begin{table}[h]
\centering
\small
\renewcommand{\arraystretch}{1.8}
\resizebox{\textwidth}{!}{%
\begin{tabular}{l|c||cc|cc|cc|cc|cc|cc}
\toprule
\multirow{3}{*}{\textbf{Model}} & 
\multirow{3}{*}{\textbf{Framework}} &
\multicolumn{4}{c|}{\textbf{Safetensors}} &
\multicolumn{4}{c|}{\textbf{\projectname{}-Unencrypted}} &
\multicolumn{4}{c}{\textbf{\projectname{}-Encrypted}} \\
\cline{3-14}
& & 
\multicolumn{2}{c|}{\textbf{save\_file}} &
\multicolumn{2}{c|}{\textbf{save\_bytes}} &
\multicolumn{2}{c|}{\textbf{save\_file}} &
\multicolumn{2}{c|}{\textbf{save\_bytes}} &
\multicolumn{2}{c|}{\textbf{save\_file}} &
\multicolumn{2}{c}{\textbf{save\_bytes}} \\
& & Time(s) & Mem(MiB) & Time(s) & Mem(MiB) & Time(s) & Mem(MiB) & Time(s) & Mem(MiB) & Time(s) & Mem(MiB) & Time(s) & Mem(MiB) \\
\midrule
\midrule
%%%%%%%%%%%% 0.6B %%%%%%%%%%%%%%%%
\multirow{2}{*}{\textbf{Qwen3-0.6B}} & PyTorch 
& 2.091 & 3280.10 & 2.898 & 6072.52 
& 2.143 (+2.5\%) & 3283.92 (+0.1\%) & 2.765 (-4.6\%) & 6074.54 (+0.0\%) 
& 3.365 (+60.9\%) & 4718.428 (+43.9\%) & 4.427 (+52.7\%) & 6177.732 (+1.7\%) \\

& Numpy 
& 1.463 & 2907.52 & 2.158 & 5676.67 
& 1.529 (+4.5\%) & 2910.35 (+0.1\%) & 2.048 (-5.1\%) & 5706.90 (+0.5\%) 
& 2.464 (+68.4\%) & 4345.630 (+49.5\%) & 3.266 (+51.4\%) & 5729.865 (+0.9\%) \\
\hline
%%%%%%%%%%%% 1.7B %%%%%%%%%%%%%%%%
\multirow{2}{*}{\textbf{Qwen3-1.7B}} & PyTorch 
& 5.148 & 8192.23 & 8.460 & 15567.75
& 5.441 (+5.7\%) & 8170.89 (-0.3\%) & 8.181 (-3.3\%) & 15772.48 (+1.3\%)
& 8.790 (+70.8\%) & 12047.131 (+47.1\%) & 11.100 (+31.3\%) & 16590.942 (+6.6\%) \\

& Numpy 
& 3.667 & 7753.20 & 6.416 & 15387.19
& 3.803 (+3.7\%) & 7794.65 (+0.5\%) & 6.268 (-2.3\%) & 15402.20 (+0.1\%)
& 5.823 (58.8\%) & 11771.978 (+51.8\%) & 9.662 (50.6\%) & 15448.435 (+0.4\%) \\

\hline
%%%%%%%%%%%% 4B %%%%%%%%%%%%%%%%
\multirow{2}{*}{\textbf{Qwen3-4B}} & PyTorch 
& 11.475 & 15748.45 & 15.910 & 30897.76
& 11.659 (+1.6\%) & 15767.45 (+0.1\%) & 16.578 (+4.2\%) & 30853.16 (-0.1\%)
& 18.348 (+59.9\%) & 23441.948 (+48.9\%) & 21.549 (+35.4\%) & 31699.034 (+2.6\%) \\

& Numpy 
& 10.126 & 15355.23 & 15.915 & 30498.97
& 9.539 (-5.8\%) & 15371.31 (+0.1\%) & 15.485 (-2.7\%) & 30544.33 (+0.2\%)
& 15.786 (+55.9\%) & 23059.512 (+50.2\%) & 25.474 (+60.0\%) & 30598.595 (+0.3\%) \\

\hline
%%%%%%%%%%%% 8B %%%%%%%%%%%%%%%%
\multirow{2}{*}{\textbf{Qwen3-8B}} & PyTorch 
& 24.165 & 31695.31 & 31.289 & 62533.20
& 24.866 (+2.9\%) & 33159.83 (+4.6\%) & 31.070 (-0.7\%) & 63361.32 (+1.3\%)
& 37.690 (+56.0\%) & 47201.215 (+48.9\%) & 49.226 (+57.3\%) & 63014.935 (+0.8\%) \\

& Numpy 
& 22.027 & 31387.75 & 29.298 & 62261.16
& 22.446 (+1.9\%) & 31332.62 (-0.2\%) & 28.712 (-2.0\%) & 62182.33 (-0.1\%)
& 35.175 (+59.7\%) & 46919.301 (+49.5\%) & 43.135 (+47.2\%) & 62274.458 (+0.0\%) \\
\hline
%%%%%%%%%%%% 14B %%%%%%%%%%%%%%%%
\multirow{2}{*}{\textbf{Qwen3-14B}} & PyTorch 
& 36.291 & 56881.88 & 51.157 & 113171.01
& 35.601 (-1.9\%) & 59909.980(+5.3\%) & 54.380 (+6.3\%) & 114207.220(+0.9\%)
& 70.172(+93.4\%) & 84948.259(+49.3\%) & 84.644(+65.5\%) & 113200.304(+0.0\%) \\

& Numpy 
& 34.620 & 56581.06 & 49.080 & 112144.06
& 35.693 (+3.1\%) & 56668.190(+0.2\%) & 49.620 (+1.1\%) & 112452.750(+0.3\%)
& 62.777(+81.3\%) & 84561.434(+49.5\%) & 80.169(+63.3\%) & 112390.934(+0.2\%) \\
\hline
%%%%%%%%%%%% 32B %%%%%%%%%%%%%%%%
\multirow{2}{*}{\textbf{Qwen3-32B}} & PyTorch 
& 76.915 & 125360.93 & 116.788 & 249339.10
& 77.069 (+0.2\%)& 125406.160(+0.0\%) & 121.576 (+4.1\%) & 249149.710(-0.1\%)
& 150.366(+95.50\%) & 187929.822(+49.9\%) & 192.901(+65.2\%) & 250190.295(+0.3\%) \\

& Numpy 
& 76.032 & 125027.64 & 112.127 & 248823.40
& 77.629 (+2.1\%) & 125040.160(+0.0\%) & 115.491 (+3.0\%) & 249533.980(+0.3\%)
& 139.113(+83.0\%) & 187538.946(+50.0\%) & 176.002(+57.0\%) & 249424.583(+0.2\%) \\
\bottomrule
\end{tabular}
}
\caption{Serialization overhead in time and peak memory across different frameworks (PyTorch/NumPy) and interfaces (\texttt{save\_file}/\texttt{save\_bytes}). \textit{Safetensors}: serialize an unencrypted model using Safetensors library. \textit{\projectname{}-Unencrypted}: serialize an unencrypted model using \projectname{} library. \textit{\projectname{}-Encrypted}: serialize a fully encrypted model with \projectname{} library. Percentage values represent the overhead relative to Safetensors.}
\label{tab:serialize_eval}
\end{table}

\noindent\textbf{\textit{Time Overhead.}}  
Table~\ref{tab:serialize_eval} reports serialization time across all model sizes and frameworks. Values in parentheses denote the relative overhead compared to the Safetensors baseline. When the encryption feature is turned off, \projectname{} library introduces minimal overhead-typically under 5\% and never exceeding 10\%. While for encrypted models, we observe the time overhead of 61.3\% for the NumPy framework and 62.0\% for the PyTorch framework.
To better understand this overhead, we profiled the serialization process and found that the majority of time is not spent on encryption itself, but on memory copying. To enable safe in-place encryption, our implementation first duplicates the tensor data using \texttt{let mut buffer = data.to\_vec()}. For instance, serializing a tensor of shape \texttt{[151936, 1024]} (approximately 300MiB) takes 0.3985 seconds in plaintext, and 0.6110 seconds with encryption. Among the additional 0.2125 seconds, 0.1654 (77.8\%) seconds is attributed to the memory copy, while only 0.0458 (21.6\%) seconds are spent on actual encryption.

\noindent\textbf{\textit{Memory Footprint.}}  
The same design decisions also impact peak memory usage, as listed in Table~\ref{tab:serialize_eval}. In the case of \texttt{save\_file}, the temporary encryption buffer remains resident throughout the write process, resulting in a peak memory overhead of approximately 50\%. For \texttt{save\_bytes}, although an additional buffer is allocated to hold the serialized output, Rust's ownership model ensures that the intermediate buffer is released before the final buffer is passed to the Python runtime. As a result, peak memory consumption remains effectively unchanged for the \texttt{save\_bytes} interface.

% Although the current implementation introduces extra memory and time costs due to buffer duplication, it was intentionally designed as a minimal modification to the Safetensors codebase for rapid adoption and compatibility. A more aggressive refactor-eliminating redundant buffers-could further reduce the overhead and will be explored in future work. Even in its current form, however, \projectname{} library provides practical performance for large-scale model serialization. \huifeng{should we move this part to other place?}\yier{I suggest moving to the conclusion section.}

\begin{figure}[h]
  \centering
  \begin{minipage}[t]{0.32\textwidth}
    \centering
    \includegraphics[width=\textwidth]{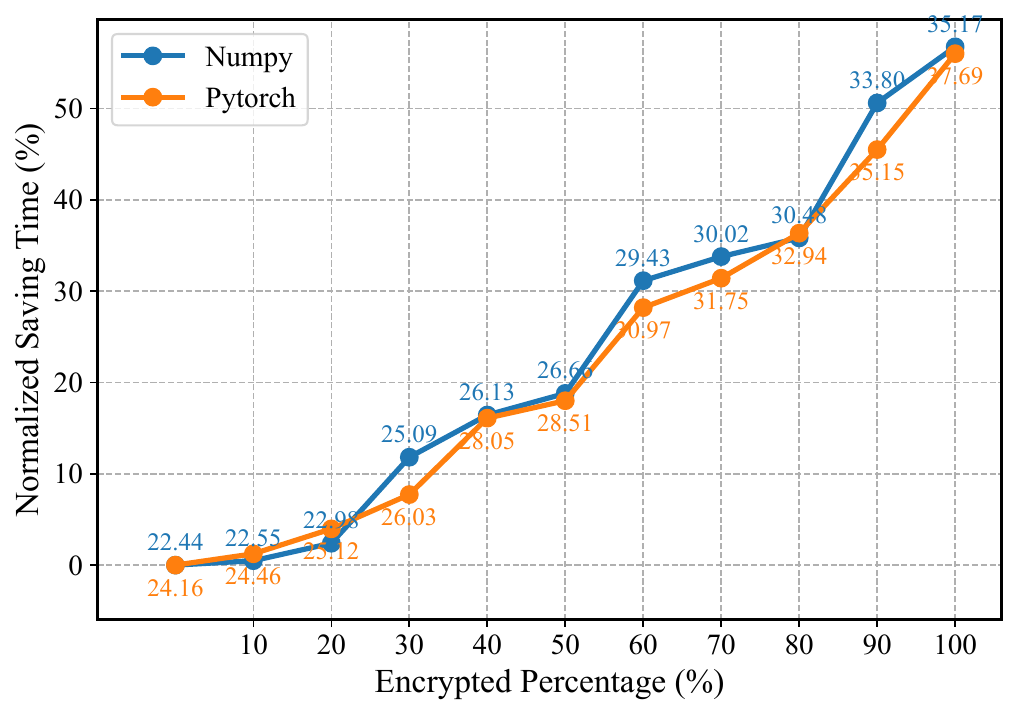}
    \captionsetup{justification=centering}
    \caption*{(a) Time with \texttt{save\_file}}
  \end{minipage}
  \hfill
  \begin{minipage}[t]{0.32\textwidth}
    \centering
    \includegraphics[width=\textwidth]{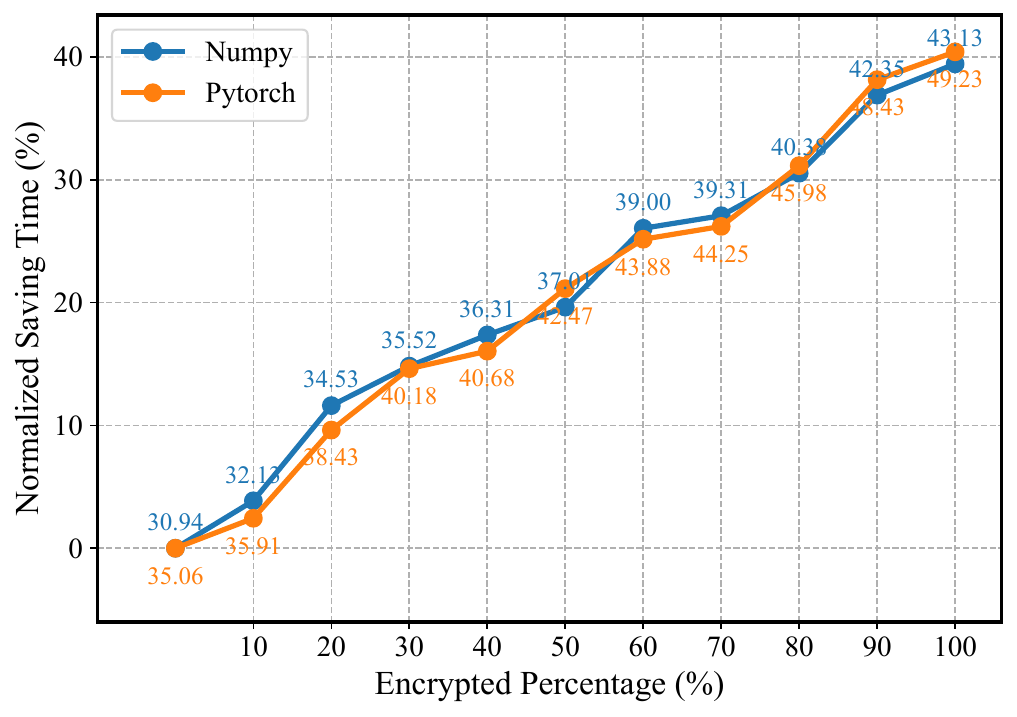}
    \captionsetup{justification=centering}
    \caption*{(b) Time with \texttt{save\_bytes}}
  \end{minipage}
  \hfill
  \begin{minipage}[t]{0.32\textwidth}
    \centering
    \includegraphics[width=\textwidth]{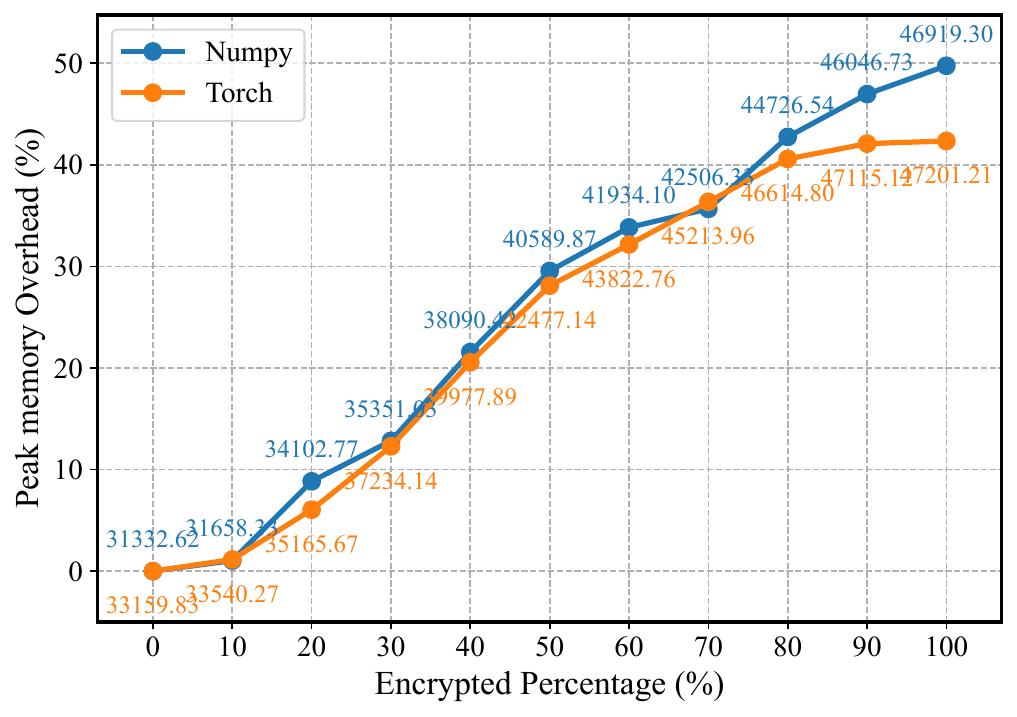}
    \captionsetup{justification=centering}
    \caption*{(c) Memory with \texttt{save\_file}}
  \end{minipage}
  \caption{Time and peak memory overhead tradeoff of saving an encrypted Qwen3-8B model using \projectname{} library as the number of encrypted tensors increases. The y-axis shows the relative overhead increase in time and peak memory compared to Safetensors. Numerical labels next to each data point indicate the absolute time (in seconds) or memory usage (in MiB).}
  \label{fig:mem-save-part}
\end{figure}

\noindent\textbf{\textit{Efficiency of Partial Encryption.}}
A key feature of \projectname{} is its support for selective encryption, enabling users to apply encryption only to sensitive parts of a model. This provides a practical balance between security and performance. Prior work in pruning and quantization suggests that not all weights contribute equally to a model’s behavior-only a small fraction of parameters (e.g., salient or super weights) carry most of the predictive power~\cite{yu2024super, lin2024awq}.
To evaluate partial encryption, we experiment on the Qwen3-8B model, which contains 399 tensors. For each encryption coverage level, we randomly select a subset of tensors to encrypt, ensuring that each level is a superset of the previous. Figure~\ref{fig:mem-save-part} (a) and (b) show how serialization time increases with the fraction of encrypted tensors. Encrypting just 10\% of the tensors incurs only about 5\% additional serialization time. Figure~\ref{fig:mem-save-part} (c) further shows that peak memory usage for \texttt{save\_file} grows approximately linearly with encryption coverage, ranging from 3\% to 45\%.
In practice, users may choose to encrypt only proprietary or fine-tuned components of the model, achieving strong confidentiality guarantees with minimal overhead.

\begin{table}[h]
\centering
\footnotesize
\renewcommand{\arraystretch}{1.3}
\resizebox{0.7\textwidth}{!}{
\begin{tabular}{lcccc}
\hline
\textbf{Model} & \textbf{Num of Tensors} & \textbf{Total Overhead} & \textbf{Header Overhead} & \textbf{Avg Overhead per Tensor} \\
\hline
Qwen3-0.6B  & 311 & 75.76 KB  & 75.76 KB  & 0.24 KB \\
Qwen3-1.7B  & 311 & 76.16 KB  & 76.16 KB  & 0.24 KB \\
Qwen3-4B    & 398 & 97.73 KB  & 97.73 KB  & 0.25 KB \\
Qwen3-8B    & 399 & 98.77 KB  & 98.77 KB  & 0.25 KB \\
Qwen3-14B   & 443 & 110.68 KB & 110.68 KB & 0.25 KB \\
Qwen3-32B   & 707 & 178.46 KB & 178.46 KB & 0.25 KB \\
\hline
\end{tabular}
}
\caption{File size overhead of saving a fully encrypted model compared to unencrypted model. 
\textit{Total Overhead}: the overall increase in file size. \textit{Header Overhead}: the overhead contributed by the header.
\textit{Avg Overhead per Tensor}: the average size increase per encrypted tensor.}
\label{tab:file_size_overhead}
\end{table}

\noindent\textbf{\textit{File Size Overhead.}}  
The file size overhead introduced by \projectname{} is directly proportional to the number of encrypted tensors. Table~\ref{tab:file_size_overhead} illustrates this trend using several models from the Qwen3 series. Since encryption does not alter the raw tensor data size, the increase in file size originates entirely from metadata added to the file header. On average, encrypting a single tensor adds approximately 0.25\,KB of header data, resulting in linear growth as encryption coverage increases. Overall, the total file size overhead remains extremely small-even for large-scale models-and is unlikely to pose any practical concern in real-world deployments.

\subsection{Overhead of Deserialization}

This subsection evaluates the runtime and memory overhead introduced by \projectname{} library during model deserialization. Both Safetensors and \projectname{} libraries provide the same interface: the \texttt{safe\_open} function is used to deserialize the model file header, while individual tensors are accessed through \texttt{get\_tensor} or \texttt{get\_slice}. In the original Safetensors design, these tensor access functions construct only lightweight tensor views and defer actual data loading due to lazy-loading behavior. In contrast, \projectname{} library decrypts the tensor payload as part of view construction, which inherently causes the data to be loaded into memory.
To ensure consistent benchmarking across configurations, we apply a simple element-wise operation (i.e., \texttt{f.get\_tensor(name).view(-1).sum()}) to each tensor during evaluation. This forces the data to be loaded in memory, thereby capturing the full time overhead and memory costs. 

\begin{table}[h]
\centering
\small
\renewcommand{\arraystretch}{1.8}
\resizebox{\textwidth}{!}{%
\begin{tabular}{l||cc|cc|cc|cc|cc|cc}
\toprule
\multirow{3}{*}{\textbf{Model}} & 
\multicolumn{4}{c|}{\textbf{Safetensors}} &
\multicolumn{4}{c|}{\textbf{\projectname{}-Unencrypted}} &
\multicolumn{4}{c}{\textbf{\projectname{}-Encrypted}} \\
\cline{2-13}
& 
\multicolumn{2}{c|}{\textbf{PyTorch}} &
\multicolumn{2}{c|}{\textbf{Numpy}} &
\multicolumn{2}{c|}{\textbf{PyTorch}} &
\multicolumn{2}{c|}{\textbf{Numpy}} &
\multicolumn{2}{c|}{\textbf{PyTorch}} &
\multicolumn{2}{c}{\textbf{Numpy}} \\
& Time(s) & Mem(MiB) & Time(s) & Mem(MiB) & Time(s) & Mem(MiB) & Time(s) & Mem(MiB) & Time(s) & Mem(MiB) & Time(s) & Mem(MiB) \\
\midrule
\midrule
%%%%%%%%%%%% 0.6B %%%%%%%%%%%%%%%%
\textbf{Qwen3-0.6B}
& 0.056 & 1776.261 & 0.945 & 2786.068
& 0.054(-3.6\%) & 1783.830(+0.4\%) & 0.946(+0.1\%) & 2730.918(-2.0\%)
 & 2.000(+3471.4\%) & 4610.102(+159.5\%) & 2.033(+115.1\%) & 4266.591(+53.1\%) \\
\hline
%%%%%%%%%%%% 1.7B %%%%%%%%%%%%%%%%
\textbf{Qwen3-1.7B}
& 0.090 & 4254.968 & 2.479 & 7558.353
& 0.094(+4.4\%) & 4262.072(+0.2\%) & 2.479(+0.0\%) & 7432.429(-1.7\%)
& 5.589(+6110.0\%) & 11988.490(+181.8\%) & 5.357(+116.1\%) & 11534.820(+52.6\%) \\
\hline
%%%%%%%%%%%% 4B %%%%%%%%%%%%%%%%
\textbf{Qwen3-4B}
& 0.157 & 8010.295 & 5.015 & 14950.114
& 0.156(-0.6\%) & 8018.251(+0.1\%) & 5.005(-0.2\%) & 14986.966(+0.2\%)
& 12.258(+7707.6\%) & 23202.010(+189.7\%) & 11.683(+133.0\%) & 22895.028(+53.1\%) \\
\hline
%%%%%%%%%%%% 8B %%%%%%%%%%%%%%%%
\textbf{Qwen3-8B}
& 0.404 & 15823.010 & 10.495 & 30763.034
& 0.421(+4.2\%) & 15837.431(+0.1\%) & 10.411(-0.8\%) & 30623.418(-0.5\%)
& 24.718(+6018.3\%) & 46984.048(+196.9\%) & 24.364(+132.1\%) & 46517.213(+51.2\%) \\
\hline
%%%%%%%%%%%% 14B %%%%%%%%%%%%%%%%
\textbf{Qwen3-14B}
& 0.766 & 28340.923 & 18.768 & 55370.795
& 0.791(+3.3\%) & 28352.791(+0.0\%) & 18.777(+0.0\%) & 55370.323(-0.0\%)
& 43.391(+5564.6\%) & 84434.608(+197.9\%) & 42.669(+127.3\%) & 83921.950(+51.6\%) \\
\hline
%%%%%%%%%%%% 32B %%%%%%%%%%%%%%%%
\textbf{Qwen3-32B}
& 1.567 & 61885.641 & 41.397 &  123035.702
& 1.621(+3.4\%) & 61935.934(+0.1\%) & 41.705(+0.7\%) & 123103.283(+0.1\%)
& 95.297(+5981.5\%) & 185506.571(+199.8\%) & 95.009(+129.5\%) & 186359.029(+51.5\%) \\
\bottomrule
\end{tabular}
}
\caption{The time and peak memory overhead of \projectname{} library during deserializing model file with different backends(PyTorch and Numpy) compared to Safetensors. \textit{Safetensors}: deserialize an unencrypted model using Safetensors. \textit{\projectname{}-Unencrypted}: use \projectname{} library to deserialize an unencrypted model. \textit{\projectname{}-Encrypted}: deserialize a fully encrypted model with \projectname{} library. Percentage values represent the increase relative to Safetensors.}
\label{tab:deserialize_eval}
\end{table}

\noindent\textbf{\textit{Time Overhead.}}
Table~\ref{tab:deserialize_eval} summarizes the total tensor loading time across different models. For unencrypted files, \projectname{} library matches the baseline Safetensors implementation, introducing negligible runtime overhead. In contrast, encrypted deserialization incurs 125.5\% (Numpy) and 5808.9\% (PyTorch) additional cost due to cryptographic processing and data movement.
To better understand the source of this overhead, we profile the loading of a representative tensor with approximately 300 MiB size. On the PyTorch framework, deserializing this tensor takes 0.4095s in total, with 0.1851s (45.2\%) spent copying the encrypted buffer, 0.0554s (13.5\%) on decryption, and 0.1683s (41.1\%) on wrapping the result into a Python-native object using \texttt{PyByteArray}. The NumPy framework shows a similar breakdown. The total deserialization takes 0.4048s, including 0.1813s (44.8\%) for buffer duplication, 0.0550s (13.6\%) for decryption, and 0.1682s (41.6\%) for NumPy array construction.
These results indicate that the dominant sources of overhead are memory copying and Python interop, rather than the cryptographic operations themselves. For completeness, we also measure header parsing latency. For example, parsing the header of a 3.72 GB model with 81 tensors requires only 0.3 ms, confirming that metadata verification is lightweight and not a performance bottleneck.

\begin{figure}[h]
  \centering
  \begin{minipage}[b]{0.49\textwidth}
    \centering
    \includegraphics[width=\textwidth]{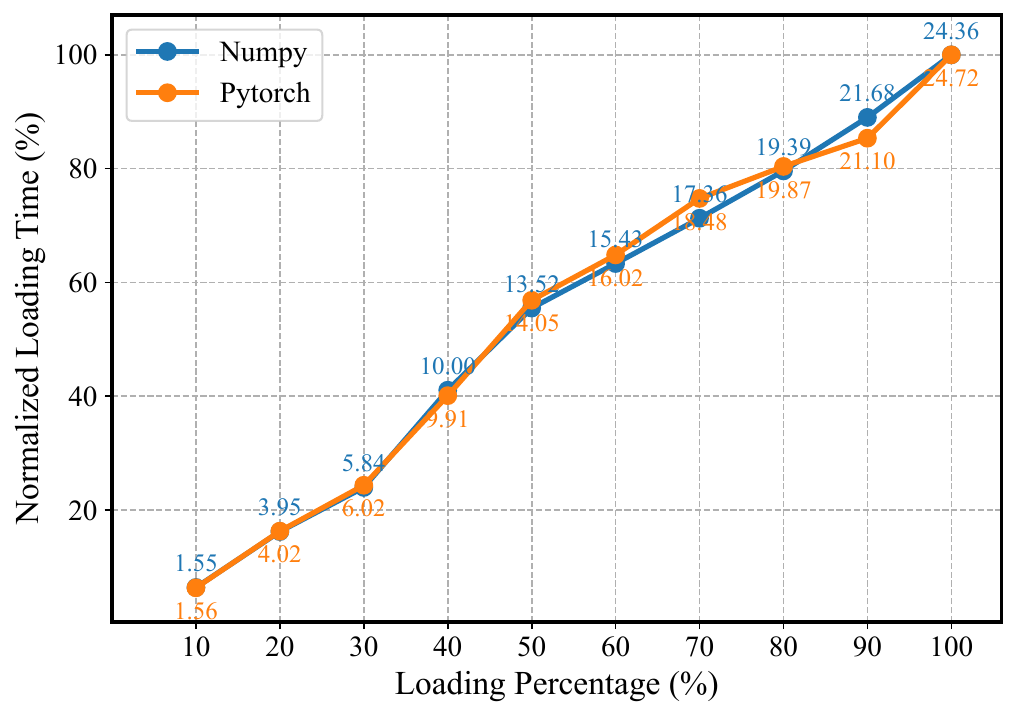}
    \captionsetup{justification=centering}
    \caption*{(a) Time}
  \end{minipage}
  \hfill
  \begin{minipage}[b]{0.49\textwidth}
    \centering
    \includegraphics[width=\textwidth]{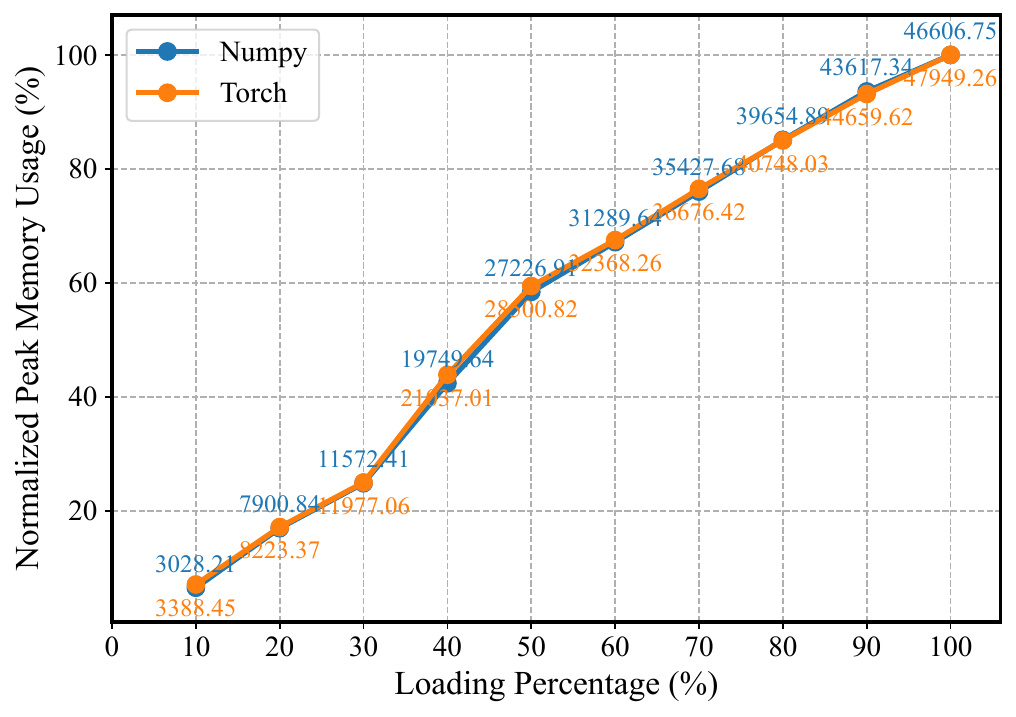}
    \captionsetup{justification=centering}
    \caption*{(b) Peak Memory Usage}
  \end{minipage}
  \caption{Loading time and peak memory usage when incrementally loading encrypted tensors from the Qwen3-8B model, reported for both NumPy and PyTorch frameworks. The y-axis shows the relative value in time and peak memory compared to loading all encrypted tensors (100\%). Each point is annotated with the corresponding absolute time (in seconds) and memory usage (in MiB).}
  \label{fig:deserialzie-part}
\end{figure}

\begin{table}[h]
\centering
\small
\renewcommand{\arraystretch}{1.3}
\resizebox{1\textwidth}{!}{
\begin{tabular}{clccccc}
\hline
\textbf{Model} & \textbf{Configuration} & \textbf{Model Load Time (s)} & \textbf{First Token Latency (s)} & \textbf{Throughput (tok/s)} & \textbf{CPU Mem (MiB)} & \textbf{Acc Mem (GiB)} \\
\hline
\multirow{3}{*}{Qwen3-0.6B} 
& Safetensors & 1.169 & 0.050 & 23.19 & 3228.30 & 1.42\\
& \projectname{}-Unencrypted & 1.160 & 0.049 & 22.59 & 3234.30 & 1.42 \\
& \projectname{}-Encrypted   & 2.713 & 0.053 & 22.03 & 4150.58 & 1.42 \\
\hline
\multirow{3}{*}{Qwen3-1.7B} 
& Safetensors & 1.760 & 0.050 & 22.43 & 6791.31 & 3.25 \\
& \projectname{}-Unencrypted & 1.699 & 0.050 & 22.77 & 5761.57 & 3.25 \\
& \projectname{}-Encrypted   & 4.747 & 0.052 & 22.47 & 7926.70 & 3.25 \\
\hline
\multirow{3}{*}{Qwen3-4B} 
& Safetensors & 2.649 & 0.050 & 21.89 & 6145.21 & 7.54 \\
& \projectname{}-Unencrypted & 2.613 & 0.048 & 22.29 & 6336.77 & 7.54 \\
& \projectname{}-Encrypted   &10.552 & 0.049 & 22.03 & 8888.01 & 7.54 \\
\hline
\multirow{3}{*}{Qwen3-8B} 
& Safetensors & 4.563 & 0.049 & 22.01 & 5415.50 & 15.31 \\
& \projectname{}-Unencrypted & 4.674 & 0.051 & 22.01 & 5401.36 & 15.31 \\
& \projectname{}-Encrypted   &25.569 & 0.049 & 22.34 & 8877.39 & 15.31 \\
\hline
\multirow{3}{*}{Qwen3-14B} 
& Safetensors & 9.558 & 0.072 & 15.63 & 5768.52 & 27.56 \\
& \projectname{}-Unencrypted & 9.388 & 0.069 & 15.80 & 5778.37 & 27.56 \\
& \projectname{}-Encrypted   &50.171 & 0.069 & 16.08 & 9341.28 & 27.56 \\
\hline
\multirow{3}{*}{Qwen3-32B} 
& Safetensors & 21.227 & 0.136 & 10.29 & 6138.77 & 61.08 \\
& \projectname{}-Unencrypted & 20.365 & 0.135 & 10.20 & 6146.98 & 61.08 \\
& \projectname{}-Encrypted   &107.714 & 0.129 & 10.36 & 9753.41 & 61.08 \\
\hline
\end{tabular}
}
\caption{Impact of integrating \projectname{} library into the \texttt{transformers} (PyTorch) workflow for model deployment and inference. 
% \textbf{Safetensors} refers to deploying an unencrypted model using the standard Safetensors. \textbf{Cryptotensors-Unencrypted} and \textbf{Cryptotensors-Encrypted} refer to deploying unencrypted and encrypted models, respectively, using \projectname{} library. 
\textit{Load Time}: the time required to load the model from files. \textit{First Token Latency}: the time from the completion of model loading to the generation of the first output token. 
 \textit{Throughput}: the average token generation rate during inference. \textit{CPU Mem} and \textit{Acc Mem}: the peak memory usage on CPU and accelerator, respectively.}
\label{tab:transformers}
\end{table}

\begin{table}[h]
\centering
\small
\renewcommand{\arraystretch}{1.3}
\resizebox{1\textwidth}{!}{
\begin{tabular}{clccccc}
\hline
\textbf{Model} & \textbf{Configuration} & \textbf{Model Load Time (s)} & \textbf{First Token Latency (s)} & \textbf{Throughput (tok/s)} & \textbf{CPU Mem (MiB)} & \textbf{Acc Mem (GiB)} \\
\hline
\multirow{3}{*}{Qwen3-0.6B}
& Safetensors     & 0.336 & 90.512 & 197.15 & 10557.58 & 1.17 \\
& \projectname{}-Unencrypted    & 0.340 & 91.100 & 195.30 & 10558.89 & 1.17 \\
& \projectname{}-Encrypted      & 2.073 & 91.062 & 193.92 & 10613.48 & 1.17 \\
\hline
\multirow{3}{*}{Qwen3-1.7B}
& Safetensors     & 0.540 & 101.911 & 198.45 & 10547.66 & 3.25 \\
& \projectname{}-Unencrypted    & 0.594 & 101.727 & 190.90 & 10560.64 & 3.25 \\
& \projectname{}-Encrypted      & 5.650 & 102.108 & 195.19 & 10666.21 & 3.25 \\
\hline
\multirow{3}{*}{Qwen3-4B}
& Safetensors     & 1.101 & 123.277 & 126.13 & 11075.59 & 7.58 \\
& \projectname{}-Unencrypted    & 1.104 & 123.406 & 125.23 & 11088.67 & 7.58 \\
& \projectname{}-Encrypted      &11.918 & 123.467 & 126.84 & 11608.82 & 7.58 \\
\hline
\multirow{3}{*}{Qwen3-8B}
& Safetensors     & 1.732 & 122.039 & 115.87 & 11095.98 & 15.30 \\
& \projectname{}-Unencrypted    & 1.806 & 122.563 & 115.61 & 11108.98 & 15.30 \\
& \projectname{}-Encrypted      &27.056 & 123.134 & 112.85 & 11233.99 & 15.30 \\
\hline
\multirow{3}{*}{Qwen3-14B}
& Safetensors     & 3.082 & 134.347 & 78.02 & 11378.91 & 27.78 \\
& \projectname{}-Unencrypted    & 3.165 & 134.765 & 76.64 & 11387.66 & 27.78 \\
& \projectname{}-Encrypted      &61.879 & 136.105 & 78.61 & 11425.65 & 27.78 \\
\hline
\multirow{3}{*}{Qwen3-32B}
& Safetensors     &7.641 & 218.374 & 43.09 & 12752.61 & 61.57 \\
& \projectname{}-Unencrypted    &7.668 & 217.488 & 42.15 & 12757.07 & 61.57 \\
& \projectname{}-Encrypted      &111.850 &218.584 & 42.93 & 12795.65 & 61.57 \\
\hline
\end{tabular}
}
\caption{Impact of integrating \projectname{} library into the \texttt{vLLM} workflow for model deployment and inference. 
% \textbf{Safetensors} refers to deploying an unencrypted model using the standard Safetensors. \textbf{Cryptotensors-Unencrypted} and \textbf{Cryptotensors-Encrypted} refer to deploying unencrypted and encrypted models, respectively, using \projectname{} library. 
\textbf{Load Time}: the time required to load the model from files. \textit{First Token Latency}: the time from the completion of model loading to the generation of the first output token. 
 \textbf{Throughput}: the average token generation rate during inference. \textit{CPU Mem} and \textit{Acc Mem}: the peak memory usage on CPU and accelerator, respectively.}
\label{tab:vllm_performance}
\end{table}

\noindent\textbf{\textit{Memory Footprint.}}
Memory usage during deserialization generally mirrors the runtime trends. For unencrypted models, \projectname{} library exhibits no noticeable difference in peak memory consumption compared to the baseline Safetensors implementation. However, encrypted models incur additional overhead due to temporary buffer allocations required during the decryption process.
As shown in Table~\ref{tab:deserialize_eval}, deserializing encrypted tensors increases peak memory usage by approximately 52.2\% on the NumPy framework and around 187.6\% on PyTorch. This overhead primarily arises from three sources. 
First, the tensor is loaded from disk into a buffer-even for unencrypted data-since our implementation does not differentiate between encrypted and unencrypted tensors for simplicity. 
Second, if the tensor is encrypted, it is decrypted into a separate buffer. 
Third, the data-either the original tensor or the decrypted buffer-is copied once more when transferred from Rust to Python, as a \texttt{np.array} or \texttt{PyBytes} object. For example, deserializing a tensor with approximately 300 MiB size results in a peak memory footprint of nearly 900 MiB during the transfer phase. As a result, for the PyTorch backend, the peak memory usage of \projectname{} under encryption is approximately 3$\times$ that of the baseline. For the NumPy backend, given that the baseline already maintains buffers for both memory mapping and the final Python object, the \projectname{} library introduces roughly 50\% additional memory overhead during deserialization on encrypted model.

\noindent\textbf{\textit{Efficiency of Lazy Decryption.}}
\projectname{} adopts a lazy decryption strategy to reduce runtime and memory overhead during deserialization. This mechanism has two complementary properties.
First, decryption is performed only for tensors that are actually accessed, avoiding global decryption. Figure~\ref{fig:deserialzie-part} (a) shows that the loading time gradually increases as more tensors are loaded from the encrypted model. Figure~\ref{fig:deserialzie-part} (b) shows that memory usage scales gradually as more encrypted tensors are accessed, demonstrating that \projectname{} library preserves on-demand loading behavior even with encryption.
Second, each tensor is decrypted at most once. Subsequent accesses reuse the previously decrypted buffer without reapplying cryptographic operations. For example, accessing a \texttt{[151936, 1024]} tensor takes 0.375s (NumPy) or 0.458s (PyTorch) on the first access, but only 0.091s and 0.193s, respectively, on the second access. In contrast, using the \texttt{Safetensors}, the same tensor is accessed in 0.110s (NumPy) or 0.043s (PyTorch) on the first access, and 0.096s and 0.038s, respectively, on the second access.
Peak memory similarly increases only during the first access, confirming that decrypted data is cached for reuse.

\subsection{End-to-End Performance Comparison}

To evaluate the real-world impact of \projectname{} library, we conduct end-to-end inference experiments by replacing the original Safetensors library with our implementation and benchmarking two widely adopted inference engines, Huggingface Transformers and vLLM.
% We replace the original Safetensors library with the \projectname{} implementation, as shown in Listing~\ref{lst:cryptotensors-migrate}. 
All experiments are conducted within Docker containers to ensure consistent and isolated runtime environments, 
utilizing accelerators for all computations.
% using 4 NVIDIA A100 40G GPUs.
For Transformers, we measure the accelerator's memory usage with Pytorch interface \texttt{max\_memory\_allocated()} and measure the throughput by timing the generation of 100 tokens.
For vLLM, metrics such as model load time, accelerator's memory usage, and throughput are directly obtained from vLLM's internal logging. While the first-token latency is measured manually using \texttt{time}.

\noindent\textbf{\textit{Performance in Transformers.}}
Table~\ref{tab:transformers} summarizes the results under the Transformers with the PyTorch framework. \projectname{} library in unencrypted mode exhibits no noticeable overhead compared to Safetensors. In encrypted mode, model loading time increases moderately, typically by 2–5$\times$, depending on model size. Crucially, this does not affect runtime efficiency. Both first-token latency and throughput remain within $\pm 6\%$ of the baseline, indicating negligible impact on inference performance.
In terms of memory usage, the accelerator's memory consumption remains consistent across all formats, confirming that decryption does not affect accelerator utilization. However, CPU memory usage increases by approximately 40\% in the encrypted case, primarily due to in-memory decryption buffers and key management overhead, as discussed previously.

\noindent\textbf{\textit{Performance in vLLM.}}
Table~\ref{tab:vllm_performance} presents the performance under the vLLM inference engine. Compared to Safetensors and the unencrypted variant, \projectname{}-enc incurs substantially higher model loading time. The overhead scales with model size-for instance, loading time increases from 3.5s (Qwen3-0.6B) to nearly 100s (Qwen3-32B), with an average slowdown of approximately 8$\times$. Since model loading is a one-time cost, steady-state performance is unaffected. First-token latency and throughput remain stable, with deviations within $\pm 7\%$ and $\pm 3\%$, respectively.
As with Transformers, the accelerator's memory usage shows no difference across configurations. CPU memory overhead from encryption remains modest, averaging less than 3\% compared to the baseline.

% compare: safetensor, cryptotensor-unencrypted, cryptotensor-encrypted, cryptotensor-partial encrypted
% compare: transformers ( framework="pt", use pytorch framework, use pytorch storage framework), transformers ( framework="tf", use tensorflow framework, use numpy storage framework ), vllm (use pytorch framework, use pytorch storage framework)

% use containers for consistent results

% we evaluate the inference performance of frameworks with cryptotensors. We evaluate the latency of the first token, the throughput, and the peak memory usage. 

% \subsection{Security Assessment}

% \huifeng{Fuzz}

% \huifeng{Attack}

% \section{Representative Deployment Scenario}
% \input{section/case_study}

% \section{Related Work}
% \input{section/related_work}

% \section{Discussion}

\section{Conclusion and Future Work}
% \huifeng{Need to make it concise} In this paper, we present \projectname{}, a secure, format-compatible extension of Safetensors designed to meet the growing demand for confidential distribution and controlled use of proprietary LLMs. Unlike prior approaches that require invasive system-level modifications or private infrastructure, \projectname{} introduces protection directly at the file format layer-embedding tensor-level encryption, cryptographic metadata, and access control policies without sacrificing compatibility with the existing model ecosystem.
% \projectname{} preserves key features of Safetensors, including lazy loading and partial deserialization, while enabling transparent key retrieval and decryption. It supports selective encryption and policy-driven access, reducing the attack surface and simplifying deployment across cloud and on-premise environments. The format is self-contained, allowing policy enforcement and secure key management to be performed using only the embedded metadata and runtime signals, with no reliance on external databases.

In this paper, we present \projectname{}, a secure, format-compatible extension of Safetensors for confidential distribution and controlled use of proprietary LLMs. Unlike prior approaches that require invasive system-level modifications or private infrastructure, \projectname{} introduces protection at the file format layer-embedding tensor-level encryption and access control policies without sacrificing compatibility with the existing model ecosystem. \projectname{} retains key Safetensors features such as lazy loading and partial deserialization, and adds transparent key retrieval, selective encryption, and policy-driven access. Its self-contained design reduces the attack surface, simplifies deployment across cloud and on-premise settings, and enables secure key management using only embedded metadata and runtime signals, without external dependencies.

To prove the effectiveness of \projectname{}, we developed a proof-of-concept \projectname{} library and conducted extensive experiments. The results show that the unencrypted mode incurs negligible overhead compared to Safetensors, while the encrypted mode introduces moderate overhead in serialization and loading time. Although the current implementation introduces some overhead due to buffer duplication, it was designed as a minimal modification to the Safetensors codebase for rapid adoption and compatibility. Future work will explore more aggressive optimizations, such as eliminating redundant buffers, to further reduce this overhead.

In the future, we plan to demonstrate how CryptoTensors can be integrated into a comprehensive system for the secure deployment and usage of large models. This includes detailing how CryptoTensors can interact with remote KBS and policy enforcement mechanisms to enhance security during model inference. Additionally, we aim to extend our optimizations to improve both the speed and memory efficiency of encrypted tensor deserialization, and explore further applications in privacy-preserving serving environments. Our goal is to provide an end-to-end solution for confidential model deployment that seamlessly integrates into existing machine learning workflows.

%\projectname{} bridges the gap between academic proposals and industrial practices, offering a lightweight, scalable, and deployable solution for protecting LLMs in open and commercial environments. We believe it lays a foundation for secure model marketplaces and trustworthy AI infrastructure, supporting encrypted sharing, controlled usage, and auditability by design.

\newpage
\bibliography{ref}

\end{CJK*}
\end{document}